\documentclass[runningheads]{llncs}

\usepackage{listings, xcolor}

\definecolor{verylightgray}{rgb}{.97,.97,.97}

\lstdefinelanguage{Solidity}{
	keywords=[1]{anonymous, assembly, assert, balance, break, call, callcode, case, catch, class, constant, continue, constructor, contract, debugger, default, delegatecall, delete, do, else, emit, event, experimental, export, external, false, finally, for, function, gas, if, implements, import, in, indexed, instanceof, interface, internal, is, length, library, log0, log1, log2, log3, log4, memory, modifier, new, payable, pragma, private, protected, public, pure, push, require, return, returns, revert, selfdestruct, send, solidity, storage, struct, suicide, super, switch, then, this, throw, transfer, true, try, typeof, using, value, view, while, with, addmod, ecrecover, keccak256, mulmod, ripemd160, sha256, sha3}, 
	keywordstyle=[1]\color{blue}\bfseries,
	keywords=[2]{address, bool, byte, bytes, bytes1, bytes2, bytes3, bytes4, bytes5, bytes6, bytes7, bytes8, bytes9, bytes10, bytes11, bytes12, bytes13, bytes14, bytes15, bytes16, bytes17, bytes18, bytes19, bytes20, bytes21, bytes22, bytes23, bytes24, bytes25, bytes26, bytes27, bytes28, bytes29, bytes30, bytes31, bytes32, enum, int, int8, int16, int24, int32, int40, int48, int56, int64, int72, int80, int88, int96, int104, int112, int120, int128, int136, int144, int152, int160, int168, int176, int184, int192, int200, int208, int216, int224, int232, int240, int248, int256, mapping, string, uint, uint8, uint16, uint24, uint32, uint40, uint48, uint56, uint64, uint72, uint80, uint88, uint96, uint104, uint112, uint120, uint128, uint136, uint144, uint152, uint160, uint168, uint176, uint184, uint192, uint200, uint208, uint216, uint224, uint232, uint240, uint248, uint256, var, void, ether, finney, szabo, wei, days, hours, minutes, seconds, weeks, years},	
	keywordstyle=[2]\color{teal}\bfseries,
	keywords=[3]{block, blockhash, coinbase, difficulty, gaslimit, number, timestamp, msg, data, gas, sender, sig, value, now, tx, gasprice, origin},	
	keywordstyle=[3]\color{violet}\bfseries,
	identifierstyle=\color{black},
	sensitive=false,
	comment=[l]{//},
	morecomment=[s]{/*}{*/},
	commentstyle=\color{black}\ttfamily,
	stringstyle=\color{red}\ttfamily,
	morestring=[b]',
	morestring=[b]"
}

\usepackage{graphicx}
\usepackage{float}
\usepackage{subcaption}

\begin{document}

\title{Rectifying Administrated ERC20 Tokens}

\author{Nikolay Ivanov \and
Hanqing Guo \and
Qiben Yan}
\institute{SEIT Lab, Michigan State University, East Lansing, MI 48824, USA
\email{\{ivanovn1,guohanqi,qyan\}@msu.edu}
\authorrunning{N. Ivanov et al.}}

\maketitle

\begin{abstract}
ERC20 token is the most popular type of Ethereum smart contract. The daily transaction volume of these tokens exceeds 100 billion dollars, which agitates the popular notions of ``decentralized banking'' and ``tokenized economy''. Yet, it is a common misconception to assume that the decentralization of blockchain entails the decentralization of smart contracts deployed on this blockchain. In practice, the developers of smart contracts implement \emph{administrating patterns}, such as censoring certain users, creating or destroying balances on demand, destroying smart contracts, or injecting arbitrary code. These routines, which are designed to tightly control the operation of these smart contracts, turn an ERC20 token into an \emph{administrated token} --- the type of Ethereum smart contract that we scrutinize in this research.

We discover that many smart contracts are administrated, which means that their owners solely possess an omnipotent power over these contracts. Moreover, the owners of these tokens carry lesser social and legal responsibilities compared to the traditional centralized actors that those tokens intend to disrupt. This entails two major problems: a) the owners of the tokens have the ability to quickly steal all the funds and disappear from the market; and b) if the private key of the owner's account is stolen, all the assets might immediately turn into the property of the attacker. Therefore, the administrated ERC20 tokens are not only dissimilar to the traditional centralized asset management tools, such as banks, but  they are also more vulnerable to adversarial actions by their owners or attackers. We develop a pattern recognition framework based on 9 syntactic features characterizing administrated ERC20 tokens, which we use to analyze existing smart contracts deployed on Ethereum Mainnet. Our analysis of 84,062 unique Ethereum smart contracts reveals that nearly 58\% of them are administrated ERC20 tokens, which accounts for almost 90\% of all ERC20 tokens deployed on Ethereum.

To protect users from the frivolousness of unregulated token owners without depriving the ability of these owners to properly manage their tokens, we introduce \emph{SafelyAdministrated} --- a library that enforces a responsible ownership and management of ERC20 tokens. The library introduces three mechanisms: \emph{deferred maintenance}, \emph{board of trustees} and \emph{safe pause}. We implement and test \emph{SafelyAdministrated} in the form of Solidity abstract contract, which is ready to be used by the next generation of safely administrated ERC20 tokens.
\end{abstract}

\section{Introduction}
Millions of Ethereum smart contracts operate hundreds of billions of dollars worth of assets. ERC20 fungible token is the most popular type of smart contract in Ethereum, often compared to decentralized bank account. Ethereum has two type of accounts: externally owned accounts (EOAs) and smart contracts. An EOA has an associated private key and can deploy smart contracts, but cannot execute custom code. On the other hand, a smart contract can execute custom code, but it does not have any associated private key for determining its owner. The deploying EOA of the contract does not automatically own this smart contract, unless this functionality is manually implemented by the contract developer. Moreover, any functionality related to ownership, role-based access, or other special permissions must be manually implemented by the developer; otherwise, the contract becomes orphaned at the moment it is deployed.

\begin{figure}
    \centering
    \includegraphics[width=0.8\textwidth]{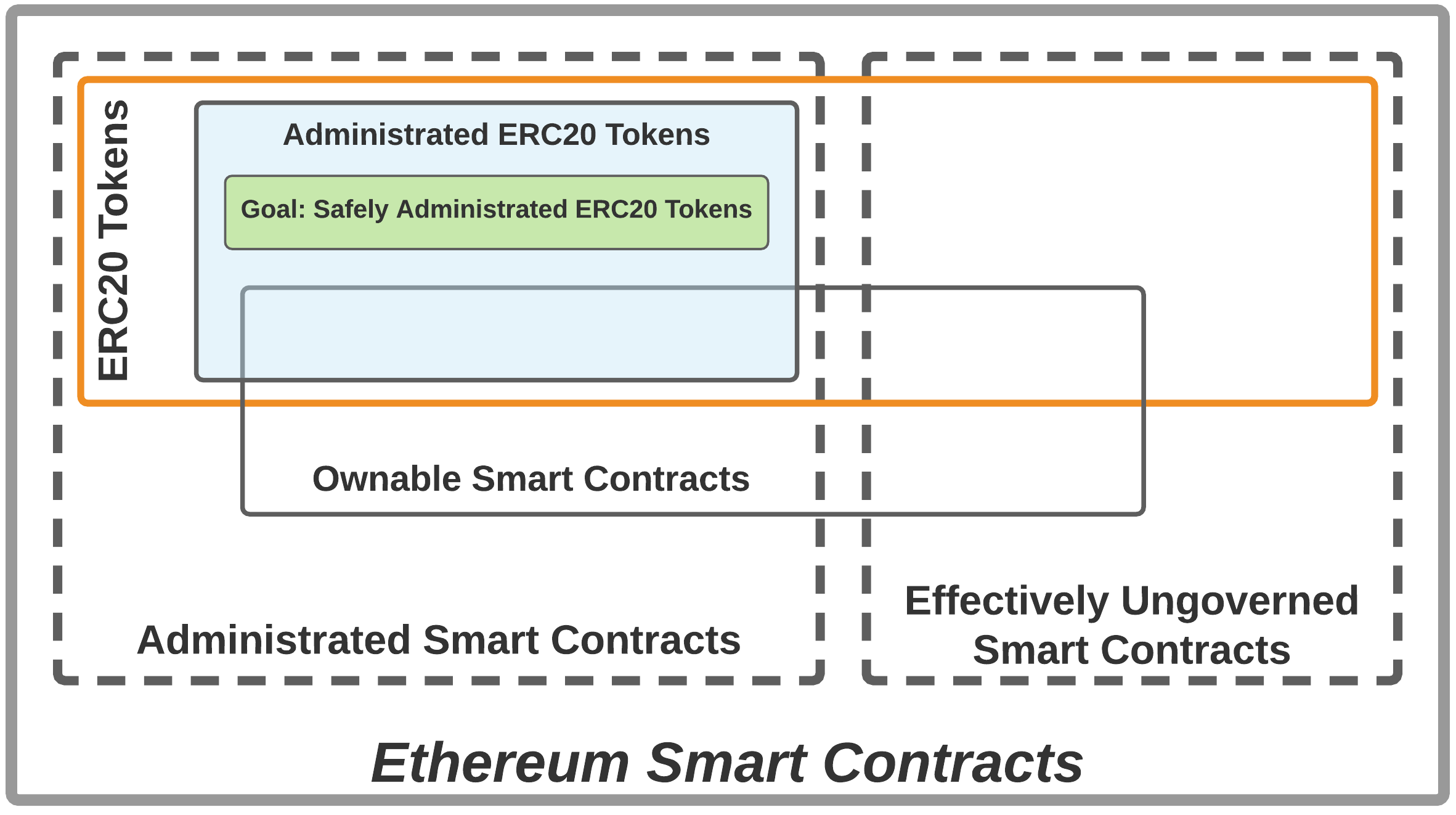}
    \caption{Venn diagram of different types of Ethereum smart contracts.}
    \label{fig:venn}
\end{figure}

Many smart contracts use routines from the OpenZeppelin Contracts~\cite{openzeppelin-contracts} library for implementing ownership and role-based access in the smart contracts. A recent analysis by Zhou et al.~\cite{zhou2020ever} shows that at least 2.1 million Ethereum smart contracts, out of 5.8 million total, use the \texttt{onlyOwner} modifier from the OpenZeppelin Contracts library, which allows only a certain user (i.e., owner) to call the functions of the smart contract implemented with this modifier. Fig.~\ref{fig:venn} shows a Venn diagram of the relationships between different subsets of Ethereum smart contracts from the perspective of this research. Specifically, we subdivide all smart contracts into two major categories: \emph{administrated contracts}, and \emph{effectively ungoverned smart contracts}, particularly emphasizing that not all contracts that have an owner are necessarily administrated, as the ownership may be  purely symbolic sometimes or only allows  harmless operations. The administrated smart contracts are characterized by two major properties: a) there is at least one Ethereum account whose owner possesses a unique privileged status; b) the privileged status allows the user to perform actions that may affect other users of the smart contract. These two properties constitute the difference between the administrated and ownable smart contracts: the ownable smart contract must only meet the first property; however, there are smart contracts that have an owner, but this owner has no power to disrupt the operation of the smart contract\footnote{The smart contracts deplyed at \texttt{0xdf4df8ee1bd1c9f01e60ee15e4c2f7643b690699} and \texttt{0x5dc60c4d5e75d22588fa17ffeb90a63e535efce0} are two (out of many) examples of ownable non-administrated contracts.}. We further refer to non-administrated smart contracts as \emph{effectively ungoverned}, the set that includes the ownable non-administrated contracts, and many of them are ERC20 tokens\footnote{A typical example of an effectively ungoverned token is the popular ChainLink Token deployed at \texttt{0x514910771AF9Ca656af840dff83E8264EcF986CA}.}. In this work, however, we zero in on the administrated ERC20 tokens, and our goal is to introduce a novel subset of these tokens --- \emph{safely administrated ERC20 tokens}.

The obvious popularity of owned smart contracts and ERC20 tokens leads us to the following research question: \textit{how many unique administrated ERC20 tokens are deployed on Ethereum?} To answer this question, we develop an extractor of 9 syntactic features characterizing administrated ERC20 tokens. We then gather 1,173,271 open source smart contracts written in Solidity programming language, and by removing the duplicates, we reduce the dataset to 84,062 unique, independent, and identically distributed (i.i.d.) smart contracts. We further select 385 random contracts for manual labeling in order to choose the most accurate classifier among several candidates. Finally, we use the 9 features and the chosen classifier to determine the approximate percentage of administrated ERC20 contracts deployed on the Ethereum Mainnet blockchain. Our evaluation shows that nearly 58\% of all the smart contracts and almost 90\% of all ERC20 tokens are administrated ERC20 tokens. To the best of our knowledge, \emph{we are the first to conduct the Ethereum-wide evaluation of administrated ERC20 tokens and quantify their ubiquity.}

To mitigate the potential adverse effects of administrated ERC20 tokens in a low-regulated economic environment, we propose \emph{SafelyAdministered} --- a Solidity library that allows developers of ERC20 tokens to implement most common administrated patterns in a safe and responsible way, thereby increasing the trust towards their products without sacrificing the need to retain control over certain operations (e.g., upgrade).

In summary, we make the following contributions:
\begin{itemize}
    \item We analyze the class of administrated ERC20 tokens and show that these contracts are more owner-controlled and less safe than the services they try to disrupt, such as banks and centralized online payment systems.
    
    \item We develop a binary classifier for identification of administrated ERC20 tokens, and conduct extensive data analysis, which reveals that nearly 9 out of 10 ERC20 tokens on Ethereum are administrated, and thereby unsafe to engage with even under the assumption of trust towards their owners.
    
    \item We design and implement \emph{SafelyAdministrated} --- a Solidity abstract class that safeguards users of administrated ERC20 tokens from adversarial attacks or frivolous behavior of the tokens' owners.
\end{itemize}

\section{Background}

\noindent \textbf{Smart Contracts and EVM.}
A smart contract is a program deployed on a blockchain and executed by the blockchain's virtual machine (VM). A smart contract consists of a set of functions that can be called through blockchain transactions. Most smart contracts are written in a high-level special-purpose programming language, such as Solidity or Vyper, and compiled into the bytecode for deployment and execution on a blockchain VM. The Ethereum Virtual Machine (EVM) is the blockchain VM for executing Ethereum smart contracts.

\noindent \textbf{Externally Owned Account.}
Ethereum blockchain has two types of accounts: smart contract account and Externally Owned Account (EOA). Both EOAs and smart contract accounts can be referenced by their 160-bit public addresses. EOAs can be used to call the functions of smart contracts via signed transactions. 

\noindent \textbf{Solidity.}
Solidity is the most popular programming language for EVM smart contract development, which syntax is similar to JavaScript and C++. The source code of a smart contract written in Solidity needs to be compiled into bytecode before being deployed on EVM. All smart contracts analyzed in this study are written in Solidity.

\noindent \textbf{ERC20 Tokens.}
ERC20 is the most popular standard for implementing fungible tokens\footnote{Each fungible token has the same value and does not possess any special characteristics compared with other tokens of the same type.} in Ethereum smart contracts. Some of the most traded alternative cryptocurrencies (altcoins) are ERC20-compatible smart contracts deployed on Ethereum Mainnet , such as ChainLink and BinanceCoin. The ERC20 standard  defines an interface with 6 mandatory functions, 2 mandatory events, and 3 optional properties that a smart contract should implement in order to become an ERC20 token to interact with ERC20-compliant clients\footnote{https://eips.ethereum.org/EIPS/eip-20}.

\noindent \textbf{OpenZeppelin Contracts.}
\emph{OpenZeppelin Contracts} is a library of smart contracts that have been extensively tested for adherence to best security practices. These smart contracts are considered to be the de-facto standardized implementations of popular smart contract code patterns~\cite{antonopoulos2018mastering}. The OpenZeppelin project provides a rich code base for ERC20 token developers~\cite{erc20-contracts}. Most ERC20 tokens, as well as the administrated patterns in these tokens, are implemented by inhereting routines from the OpenZeppelin Contracts library.

\section{Administrated ERC20 Patterns}\label{sec:patterns}
In this section, we elaborate upon five general re-centralization patterns that we observe in Ethereum smart contracts\footnote{The discovery of these patterns has been largely facilitated by a manual examination of approximately 3,800 source codes of smart contracts in the course of our previous research.}. 

\subsection{Self-destruction}
EVM opcode \texttt{SELFDESTRUCT}\footnote{This opcode is formerly known as \texttt{SUICIDE}. In this context, the word ``remove'' means that the contract is no longer available for transactions; however the entire transaction history of the contract is still retained by the blockchain.} allows to remove a smart contract from the blockchain. To provide further incentive for owners to remove unused contracts, the address supplied as an argument of \texttt{SELFDESTRUCT} call receives the entire Ether cryptocurrency balance of the smart contract. Solidity uses the built-in function \texttt{selfdestruct()} to initiate the removal of the smart contract --- if this functionality is implemented, the administrator (or an attacker impersonating the administrator) can trigger it at any moment, effectively destroying all users' assets with a single transaction. Fig.~\ref{fig:listing1} shows a real-world example of such a pattern.

\begin{figure}[t]
    \centering
\begin{lstlisting}[language=Solidity]
function kill() public onlyAdmin {
  selfdestruct(payable(msg.sender));
}
\end{lstlisting}
    \caption{A snippet of an administrated self-destruction pattern in the contract deployed at \texttt{0xbF3d14995D4A4A719A3B9101DE60baa47De60F39}.}
    \label{fig:listing1}
\end{figure}

\subsection{Deprecation}
With the exception of self-destruction, the source code of an Ethereum smart contract is immutable, which impedes the ability for developers to deliver new features or fix existing bugs. To address this limitation, some developers of smart contracts implement a bypass scheme, in which a contract can be declared as \emph{deprecated} by the owner, resulting in the redirection of the users' transactions towards functions of a new contract. The danger of this scheme stems from the fact that it grants the owner of the contract an ability to replace the code of some critical functions with arbitrary ones. Fig.~\ref{fig:listing2} shows a real-world example of the deprecation pattern.

\begin{figure}[t]
    \centering
\begin{lstlisting}[language=Solidity]
// deprecate current contract in favour of a new one
function deprecate(address _upgradedAddress) public onlyOwner {
  deprecated = true;
  upgradedAddress = _upgradedAddress;
  Deprecate(_upgradedAddress);
}
\end{lstlisting}
    \caption{A snippet of an administrated deprecation pattern in the TetherUSD smart contract deployed at \texttt{0xdAC17F958D2ee523a2206206994597C13D831ec7}, which allows the owner to effectively inject the code of the contract with an arbitrary one.}
    \label{fig:listing2}
\end{figure}

\subsection{Change of Address}
Another administration strategy is the ability for the owner of a smart contract to change certain critical addresses, such as recipients of fees or accounts associated with certain roles. As shown in our previous study~\cite{ivanov2021targeting}, a replacement of a public address in a smart contract can lead to an acquisition of the funds by the owner of the contract. Fig.~\ref{fig:listing3} demonstrates such an address changing pattern.

\begin{figure}[t]
    \centering
\begin{lstlisting}[language=Solidity]
function setFee(address to) public onlyOwner{
  fee = to;
}
\end{lstlisting}
    \caption{A snippet of a change-of-address pattern in the smart contract deployed at \texttt{0x350BDC46d931712d83ef989725Ba4904C487F360}. The exploitation of such pattern has been demonstrated in previous research.}
    \label{fig:listing3}
\end{figure}

\subsection{Change of Parameters}
Another administration pattern is characterized by the change of certain parameters by the owner, which may affect the ability by a user of the contract to perform certain operations. For example, if the owner is allowed to arbitrarily change the amount of withdrawal fees, this parameter might be set to a very large value (e.g., 99\%), effectively preventing withdrawal of funds by the user. Another example of this pattern is shown in Fig.~\ref{fig:listing4}, where the owner of the contract exercises an unbounded power to manage administrators of the smart contract.

\begin{figure}[t]
    \centering
\begin{lstlisting}[language=Solidity]
function setAdmin(address newAdmin, bool activate) onlyOwner public {
  admins[newAdmin] = activate;
}
\end{lstlisting}
    \caption{A snippet of a change-parameter pattern in the smart contract deployed at \texttt{0x18c210013ea6cbe99b2dacdc9cfcb6e07458f0ca}.}
    \label{fig:listing4}
\end{figure}

\subsection{Minting and Burning}
An increase of a token supply of an ERC20 contract is called \emph{token minting}, and the reduction of supply of tokens is called \emph{burning}. Since the entire supply of tokens is partitioned between owners in a way that there are no balances belonging to nobody, minting a token means to increase someone's balance, and burning a token means to reduce someone's balance. Although most tokens are minted or burned as a result of a certain event, such as token creation, token swap, crowdsale, or exchange into Ether balance, some contracts allow privileged users to arbitrarily mint or burn tokens, which is a dangerous action that even highly centralized commercial banks normally cannot do. Fig.~\ref{fig:listing5} demonstrates an example of the minting and burning pattern implemented in a deployed Ethereum smart contract.

\begin{figure}[t]
    \centering
\begin{lstlisting}[language=Solidity]
function mint(address account, uint amount) public onlyOwner {
  _mint(account, amount);
}
function burn(address account, uint amount) public onlyOwner {
  _burn(account, amount);
}
\end{lstlisting}
    \caption{A snippet of a minting and burning patterns in the smart contract deployed at \texttt{0x82bfdd53dd95efa2c3e92543f28d46c566bf4b8a}.}
    \label{fig:listing5}
\end{figure}

\section{Administrated Tokens in the Wild}\label{sec:study}

\begin{figure}
    \centering
    \includegraphics[width=\textwidth]{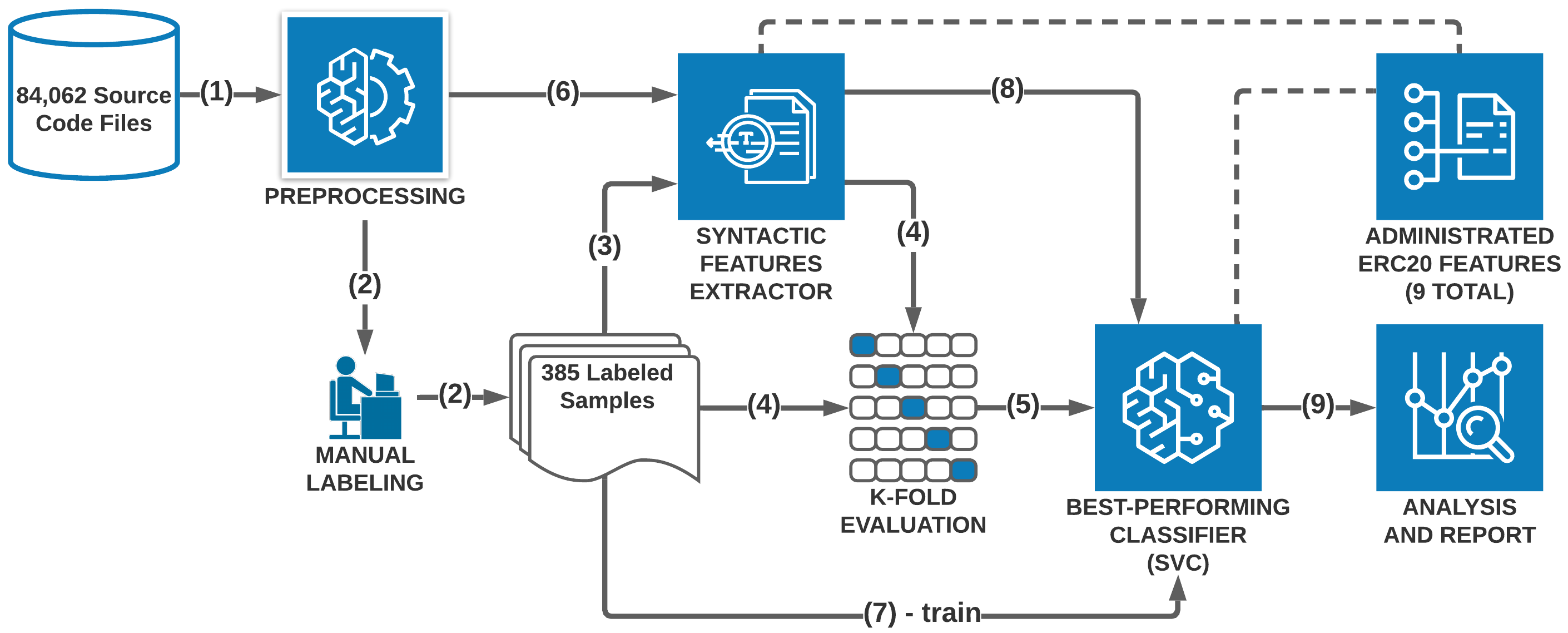}
    \caption{\textbf{General worflow of the analysis of administrated ERC20 tokens.} The workflow includes 9 major steps. \textbf{(1)}: Pre-process input samples to remove comments and parse multi-part JSON files. \textbf{(2)}: Pick 385 samples from 84,062 unique source code files and manually assign them into two classes: a) administrated ERC20 tokens, and b) others. \textbf{(3)}: Extract 385 feature vectors corresponding to the labeled samples. \textbf{(4)}: Use 385 labeled samples and the corresponding feature vectors to evaluate the performance of 9 different classifiers using the K-fold methods (with $k=5$). \textbf{(5)}: Choose the best performing classifier on the 385 labeled samples with the given 9 features. \textbf{(6)}: Extract 84,062 feature vectors corresponding to the entire data set. \textbf{(7)}: Train the classifier with the 385 labelled samples. \textbf{(8)}: Classify all the samples using the trained classifier. \textbf{(9)}: Analyze and report the results.}
    \label{fig:workflow}
\end{figure}

In this section, we use a pattern recognition method to search for administrated ERC20 tokens in the Ethereum Mainnet network, as shown in Fig.~\ref{fig:workflow}. We start the process with preprocessing all the input samples by removing comments and extracting source codes from multi-part JSON files\footnote{The smart contracts that include several files are represented as JSON arrays in our dataset. Preprocessing these arrays also includes an additional step of replacing the escaped characters, such as newlines and quotes, with their original ASCII codes.}. Then we randomly select 385 samples from 84,062 unique source code files and manually assign (label) them into two classes: a) administrated ERC20 tokens, and b) others. After that, we extract 385 9-dimensional feature vectors corresponding to the labeled samples, with the assumption that all the samples are identical and independently distributed (i.i.d). Then we use 385 labeled samples and the corresponding feature vectors to evaluate the performance of 9 different classifiers using the K-fold method (with $k=5$). Next, we choose the best performing classifier, i.e., the one that demonstrated the higher accuracy during the evaluation stage (i.e., SVC). After that, we extract 84,062 feature vectors corresponding to the entire data set. Next, we train the SVC classifier with the 385 labelled samples. Due to the i.i.d. assumption, we can now classify all the samples using the trained SVC model. Finally, we gather the output and analyze the results.

\subsection{Data Set}\label{sec:dataset}

First, we gather 1,173,271 open-source smart contracts from Etherscan\footnote{https://etherscan.io/}, and by removing duplicates (using \texttt{fdupes}\footnote{https://github.com/adrianlopezroche/fdupes}), reduce the size of the database to 84,062 distinct smart contracts. Then, we remove all comments from the data points (i.e., source code files), and select 385 random contracts for manual labelling  using the following formula:

\begin{equation}
    n = \frac{N}{1+N \cdot (1-c)^2}. 
    \label{eq:solvins}
\end{equation}

Eq.~\ref{eq:solvins} is the Slovin's formula~\cite{burt2009elementary}, which statistically determines a required representative sample size for a given data size and desired confidence level. $N$ is the original population of smart contracts, i.e., $N=84,062$, and $n$ is the sample size that we choose to represent the population. $c$ is the confidence level that represents the certainty that the sample size represents the population. We set the confidence level as $95\%$ (precisely, $94.915\%$), leading to sample size $n=385$, which can be split into two partitions of 77 and 308 samples for k-fold evaluation with $k=5$.

\subsection{ERC20 Administration Features}\label{sec:features}

Our knowledge of the administration features in ERC20 tokens stems from our experience of manual analysis of around 3,800 source codes of Ethereum smart contracts. 
The experience of manual analysis of thousands of smart contracts, which has taken more than 140 person/hours, allows us to recognize all existing administration patterns. 
As a result, we have developed 9 syntactic signatures which are intuitively well-separated and independent \emph{because we have observed various combinations of these signatures in administrated smart contracts}. This led us to designing 9 syntactic features, denoted $f_1 \ldots f_9$ that produce one of two binary values: 1 --- the corresponding syntactic signature is present; 0 --- the signature is absent. 
Below is the brief description of the syntactic signatures that the 9 features correspond to.

\subsubsection{$f_1$: ERC20 Interface Implementation.}
The goal of this research is to identify administrated ERC20 tokens. In order to separate ERC20 tokens from other types of smart contracts, feature $f_1$ extractor detects the simultaneous presence of syntactic identifiers corresponding to the eight mandatory items of the ERC20 interface, as described in the EIP-20 standard.

\subsubsection{$f_2$: Administrated Self-destruction Signature.}
If the owner of a smart contract implements a self-destruction procedure, they may remove the contract from the Ethereum ecosystem with a single transaction, simultaneously acquiring all the Ether balance of the contract. Feature $f_2$ detects such a signature, both in old versions of Solidity and the modern ones (the exact procedure differs for different versions of the language).

\subsubsection{$f_3$: Pausable Functionality Signature.}
The owner of a smart contract can inhibit any operations with the contract at their will for indefinite period of time. Although pausing a smart contract does not allow to directly acquire Ether or token balances, it may have dire consequences if the owner's private key is stolen by an attacker or lost while the token is paused. Feature $f_3$ is intended to identify signatures of such pausable tokens.

\subsubsection{$f_4$: Contract Deprecation Signature.}
Since Ethereum smart contracts are non-modifiable, the only means of upgrading the contract is to deprecate the existing contract and refer the users to the new one using inter-contract calls (ICCs). Unfortunately, this procedure allows the owner of the smart contract to effectively introduce any arbitrary code. Feature $f_4$ extracts the signatures of contract deprecation functionality, which is one of the most dangerous patterns in administrated ERC20 tokens.

\subsubsection{$f_5$: Minting and Burning Signatures.}
The ability for a privileged user to arbitrary create and remove tokens, known as minting and burning respectively, is a major concern associated with administrated ERC20 tokens. Feature $f_5$ represents the signature of a minting and/or burning in the smart contract, which execution can only be triggered by a privileged user (administrator).

\subsubsection{$f_6$: Role-restricted Transfers and Withdrawals.}
Another signature of an administrated ERC20 token is the ability for a privileged user to perform arbitrary token or Ether cryptocurrency transfers and withdrawals of the funds that do not belong to these users. Feature $f_6$ corresponds to the syntactic signature related to such transfer and withdrawal functionality under a privileged access.

\subsubsection{$f_7$: Function-disabling Modifiers.}
Some function modifiers do not directly check for the identity of privileged users; instead, they use the parameters previously changed by an administrator to decide whether the function needs to be executed. Feature $f_7$ is related to such modifiers that are capable of disabling the execution of a function based on a parameter adjustable by the contract's privileged user.

\subsubsection{$f_8$: Direct Checks of a Sender Address.}
Although modifiers are popular means of granting privileged access to certain functions of a smart contract, some administrated contracts use direct checks of the \texttt{msg.sender} or \texttt{msg.origin} values. Feature $f_8$ targets the direct (i.e., bypassing Solidity modifiers) transaction identity checks, which predominantly make sense within the administrated smart contracts context.

\subsubsection{$f_9$: Freezing, Halting, or Killing Methods.}
A list of some specific frequently occurring function names, such as ``freeze'', ``halt'', and ``kill'' empirically strongly correlate with the administrated property of ERC20 tokens. Feature $f_9$ detects the presence of such frequently used functions that almost always indicate an administration pattern.

\subsection{Classifier Evaluation and Model Selection}
We use 385 manually labeled samples to evaluate the performance of 9 popular classifiers using the K-fold method with $k=5$. Table~\ref{tab:classifiers} summarises the classification models used for evaluation and the accuracy of each of these models using the K-fold evaluation method with 385 labeled samples. The evaluation demonstrates that 8 out of 9 classifiers stay within the $95\% \ldots 97\%$ accuracy range, except for the Gaussian Naive Bayes classifier, which performance is slightly above $61\%$.

\begin{table}
    \centering
    \caption{Tested classifiers.}
    \label{tab:classifiers}
    \begin{tabular}{|l|l|c|}
    \hline
        \textbf{Model} & \textbf{Parameters} & \textbf{Accuracy} \\
    \hline
       Support Vector Classifier (SVC)  & \texttt{scikit-learn} default & 96.6233\% \\
       \hline
       Decision Tree  & $max. depth = 9$ & 96.3636\% \\
       \hline
        K-Nearest Neighbors (K-NN)  & $k=1$ & 95.5844\% \\
       \hline
       Random Forest  & \texttt{scikit-learn} default & 96.3636\% \\
       \hline
       Gaussian Naive Bayes  & \texttt{scikit-learn} default & 61.0389\% \\
       \hline
       Linear Discriminant Analysis (LDA)  & $n\_components = 1$ & 96.3636\% \\
       \hline
       Gradient Boosting  & \texttt{scikit-learn} default & 96.3636\% \\
       \hline
       Adaptive Boosting (AdaBoost)  & \texttt{scikit-learn} default & 95.0649\% \\
       \hline
       Multi-Layer Perc. Classifier (MLPC)  & $alpha=1, max\_iter=1000$ & 96.6233\% \\
       \hline   
    \end{tabular}
\end{table}

\subsection{Implementation and Evaluation of the Analysis Workflow}

We implement the extractors of all the 9 syntactic features using Python 3.8.5 and \texttt{re} regular expressions library. We implement the K-fold evaluation and dataset analysis using Python 3.8.5 with \texttt{sckit-learn} 0.24.1 and \texttt{numpy} 1.20.0 libraries. We randomly selected 385 smart contracts from the i.i.d. set of 84,062 and manually labeled them by human comprehension of the semantics of each of the smart contracts, which took approximately 40 person/hours of total effort.

\subsection{Results}
\begin{figure}
\centering
\begin{subfigure}{0.31\linewidth}
  \centering
    \includegraphics[width=\textwidth]{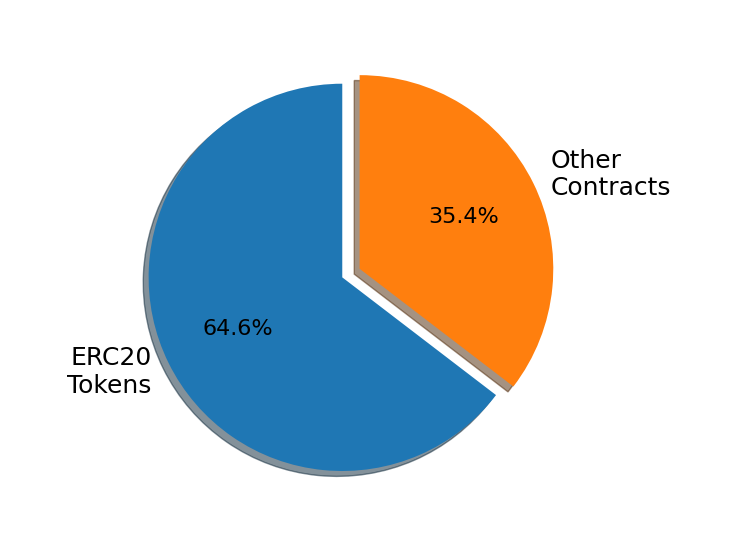}
  \caption{ERC20 tokens vs. other all other smart contracts.\\}
  \label{fig:pie1}
\end{subfigure}~~~
\begin{subfigure}{0.31\linewidth}
  \centering
    \includegraphics[width=\textwidth]{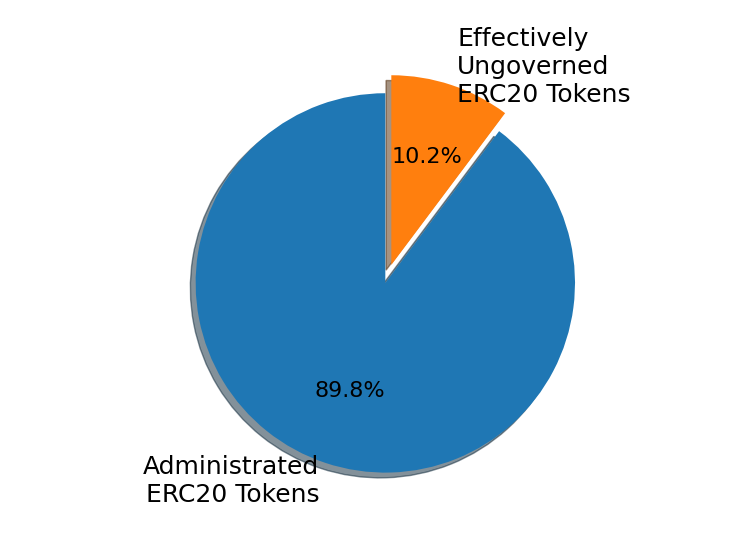}
  \caption{Administrated ERC20 tokens vs. effectively ungoverned ERC20 tokens.}
  \label{fig:pie2}
\end{subfigure}~~~
\begin{subfigure}{0.31\linewidth}
  \centering
    \includegraphics[width=\textwidth]{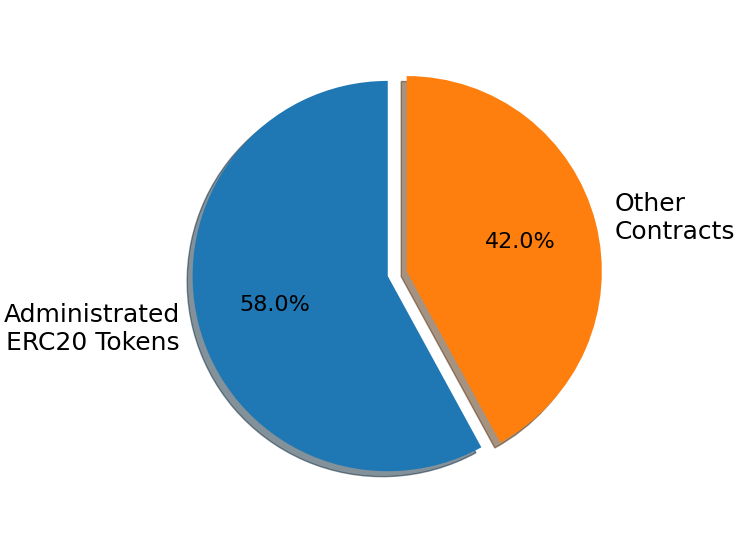}
  \caption{Administrated ERC20 tokens vs. other types of smart contracts.}
  \label{fig:pie2}
\end{subfigure}
\caption{Results of processing of 84,602 unique source codes of Ethereum smart contracts using the SVC classifier and the 9 developed syntactic features.}
\label{fig:pie}
\end{figure}

Out of 84,062 evaluated smart contracts, 54,626 have been identified as ERC20 tokens, which is around 64.6\%. As many as 39,034 contracts have been classified as administrated ERC20 tokens (by counting the occurrences of $f_1 = 1$), which is 57.96\% of all the evaluated smart contracts, and 89.76\% of all ERC20 tokens. Subsequently, only about 10\% of all ERC20 tokens are non-administrated, i.e., exhibit full decentralization and permissionless design, while the vast majority of the tokens are tightly controlled by their owners and other privileged users, effectively overriding the decentralization capability of the hosting blockchain. Fig.~\ref{fig:pie} shows the summary of the results of our analysis.

\section{SafelyAdministrated Library}\label{sec:serc20}
Existing administrated ERC20 tokens are generally unsafe because they are loosely regulated and their functionality often hinges upon a single account's private key, which can be abused by its owner or stolen by an adversary. To mitigate such an unsafe arrangement without denouncing the idea of administration or boycotting the administrated tokens,
we propose a novel solution for making these smart contracts safe. As shown in Section~\ref{sec:study}, most ERC20 tokens are administrated, and therefore potentially unsafe. However, due to their ubiquity, it would be naive to urge users to boycott 9 out of 10 of currently deployed ERC20 tokens. In this work, 
we propose a feasible ``evolutionary'' fix to the existing problem. Specifically, we realize that administrated patterns can be used by token owners without jeopardizing the safety of the contract and requiring trust from the users. For that, the current primitive administrated routines can be re-implemented to incorporate three novel concepts: \emph{deferred maintenance}, \emph{board of trustees}, and \emph{safe pausing}. The details of these three approaches are explained below.

\subsection{Deferred Maintenance}\label{sec:serc20-sliding-action-deference}
The owners of existing administrated ERC20 tokens have the ability to call the managerial functions without any announcement.
In order to prevent unannounced actions, \emph{SafelyAdministrated} library implements a mechanism of \emph{deferred maintenance}, which allows to announce the maintenance action to the users and enact it only after a certain delay. For example, if the contract is about to be upgraded, the users of the contract may be notified and decide whether they agree on the upgrade or not. If the users disagree with the upgrade, they may safely quit (i.e., sell or transfer their tokens) before the action takes into effect.

\subsection{Contract Board of Trustees}\label{sec:overlay-consensus}
In most administrated smart contracts, the privileged user (administrator) has a sole power to perform critical actions upon the smart contract, which incurs the need of trust from the users of the contract. Moreover, if the private key of the smart contract's administrator is stolen, the attacker \emph{becomes} the administrator of the contract. Essentially, the safety of the contract often hinges on a single private key belonging to a single person, which is the major concern about the administrated smart contracts. The \emph{contract board of trustees} allows to split the administrative power among multiple private keys possessed by different parties, such that the maintenance actions are only possible through a voting consensus with a pre-determined threshold.

\subsection{Safe Pause}\label{sec:safe-pause}
The ability to pause the execution of transactions in a smart contract is not necessarily a whimsical action of the contract administrator. For example, this may be a necessary action upon discovery of a zero-day vulnerability --- by pausing transactions, the administrator of the contract may prevent an exploitation of such vulnerability. However, indefinite pause may also be abused by the contract administrator, or it can be triggered by an adversary who stole a private key of the administrator's account. To prevent the adverse effects of the pause functionality, in this work we introduce a \emph{safe pause} routine, which allows to freeze all transactions in the smart contract with a forced un-freeze after a certain deadline. Moreover, once the contract is un-frozen, it cannot be frozen again for some time. This way, any of the trustees of the contract can enact an emergency pause, but no one is able to keep the contract paused indefinitely.

\subsection{Implementation}
We implement \emph{SafelyAdministrated} as an abstract Solidity class, which includes 6 functions, 3 modifiers, and 5 events, summarized in Table~\ref{tab:sa-summary}. We implemented a testing ERC20 token that inherits the \emph{SafelyAdministrated} contract, compiled it using Solc 0.8.1, and thoroughly tested its functionality to confirm that \emph{SafelyAdministrated} allows an ERC20 token to be administrated in a safe manner.

\begin{table}
    \centering
    \caption{Inheritable interfaces of \emph{SafelyAdministrated} abstract class.}
    \label{tab:sa-summary}
    \begin{tabular}{|l|c|l|}
    \hline
    \textbf{Inheritable Interface} & \textbf{Type} & \textbf{Description} \\
    \hline
    \texttt{actionCleared} & function & Check if a given action can be performed \\
    \hline
    \texttt{safelyPaused} & function & Check if contract is paused \\
    \hline
    \texttt{safelyUnpaused} & function & Check if contract is unpaused \\
    \hline
    \texttt{safelyPause} & function & Safely pause the smart contract \\
    \hline
    \texttt{safelyUnpause} & function & Safely un-pause the smart contract \\
    \hline
    \texttt{whenSafelyPaused} & modifier & Check if contract is paused \\
    \hline
    \texttt{whenSafelyUnpaused} & modifier & Check if contract is un-paused \\
    \hline
    \texttt{trusteeVote} & function & Cast trustee vote for an action \\
    \hline
    \texttt{SafelyPaused} & event & A trustee paused the contract \\
    \hline
    \texttt{SafelyUnpaused} & event & A trustee un-paused the contract \\
    \hline
    \texttt{TrusteeVoted} & event & A trustee voted for an action \\
    \hline
    ActionCleared & event & Next vote will activate the action \\
    \hline
    \texttt{ActionActivated} & event & A trustee vote activated a cleared action \\
    \hline
    \texttt{trusteeAction[0..9]} & modifier & Modifiers for nine functions subject to approval \\
    \hline
    \end{tabular}
\end{table}

\subsection{Limitation}\label{sec:discussion}
One limitation of \emph{SafelyAdministrated} is that the trustee whose vote attains the voting threshold effectively pays fees for the execution of the maintenance transaction, while other trustees pay only for execution of recording of their vote. Although we assume that this unfairness is unlikely to be important in most cases, we leave the implementation of fee reimbursement for future work.

\section{Related Work}\label{sec:relatedwork}
Currently, the major concern about the safety of smart contracts comes from security vulnerabilities in them. Researchers have proposed automated tools for detecting known smart contract vulnerabilities. Some notable security scanners for Ethereum include Oyente~\cite{luu2016making}, Mythril~\cite{mythril-github}, and Vandal~\cite{brent2018vandal}. Tsankov et al.~\cite{tsankov2018securify} propose Securify, a tool that analyzes the bytecode of Ethereum smart contracts to detect patterns associated with known security vulnerabilities. Torres et al.~\cite{torres2019art} present a taxonomy of smart contract \emph{honeypots}, which are deceptive smart contracts targeting users who attempt to exploit known vulnerabilities of smart contracts. Recently, Chen et al. propose TokenScope~\cite{chen2019tokenscope}, an automated tool, which detects the discrepancies between syntax and semantics in the functions of ERC20 tokens. In this work, we reach beyond the security vulnerabilities and explore a generally overlooked safety issue in smart contracts, i.e., administrated patterns that allow owners of ERC20 tokens (or adversaries who steal the owner's account private key) to cause a mass damage to the token owners.

The influence of private actors on blockchain resources has been a subject of concern for many years. Raman et al.~\cite{raman2019challenges} conduct a case study of decentralized web applications and identify a prevalence of \emph{re-centralization} of such apps. Griffin et al.~\cite{griffin2020bitcoin} discover that TetherUSD ERC20 token has been used for manipulating the price of cryptocurrencies. In this work, we expand the discussion about the re-centralization and private manipulation of the services that are intended to be centralized to embrace the realm of ERC20 tokens.

The public trust towards administrated ERC20 tokens may be indicative of a well-studied irrational or semi-rational human behavior. In our previous research~\cite{ivanov2021targeting}, we explore social engineering attacks in Ethereum smart contracts by demonstrating how visual cognitive bias and confirmation bias lead a user into engaging with a malicious smart contract. Fenu et al.~\cite{fenu2018ico} demonstrate the irrational behavior exhibited by many people when engaging with high-risk smart contracts involved in intitial coin offerings (ICOs). In this work, we scrutinize a new facet of semi-rational human behavior: the false assumption that most smart contracts are decentralized, permissionless, and ungoverned \emph{just because} they are deployed on a blockchain that holds these properties.

Previous studies proposed smart contract-level multi-signature voting schemes. {\AE}GIS~\cite{ferreira2020aegis} implements a voting-based mechanism, in which trusted experts vote for a security patch. Unfortunately, the voting mechanism in {\AE}GIS has been design for different context and cannot be applied, even with modifications, to the trustee-based contract maintenance scenarios. Christodoulou~\cite{christodoulou2020decentralized} introduces a decentralized voting scheme similar to the Board of Trustees used in this work. However, all the above solutions are domain-specific, and cannot be directly used for general cases, as we see it in the \emph{SafelyAdministrated} library.

\section{Conclusion}\label{sec:conclusion}
Unlike banks and other financial institutions, smart contracts are weakly regulated or unregulated at all. Simultaneously, an ERC20 token is often owned by a single account, the security of which hinges on a single private key. At the same time, we observe that market capitalization of some tokens, such as USDT and BNB, reaches billions of dollars, which means that if the administrator's private key is stolen or abused, all the funds from all users in the contract might be stolen immediately. ERC20 fungible tokens have been a hope for the next-generation tokenized economy. However, in this research we demonstrate that approximately 9 out of 10 ERC20 tokens are administrated assets that are generally less secure than traditional financial institutions and accounts. Instead of stigmatizing the widespread administration of the tokens, we deliver a solution for the honest token owners to achieve their goals in a way that is safe for both them and the users --- through implementing the novel contract ownership mechanism, which effectively prevents a single point of security failure and enforces prior notice of maintenance. At the time of writing, there is no affiliation or sponsorship, current or arranged, between the authors of this work and any banks, online payment systems, and smart contract developers mentioned or implied in this research.

\section*{Acknowledgement}\label{sec:acknowledgement}
We would like to thank Dr. Arun Ross and other anonymous reviewers for providing valuable feedback on our work. 
\clearpage
\bibliographystyle{splncs04}

\end{document}